**The Role of Distributional Overlap on the Precision Gain of Bounds for Generalization**

*Forthcoming in the American Journal of Evaluation*

Wendy Chan, Ph.D.
Assistant Professor of Education
Human Development and Quantitative Methods, Graduate School of Education
University of Pennsylvania
3700 Walnut Street
Philadelphia, PA 19104 – 6216

**Acknowledgements:**

This study was funded by the Spencer Foundation through a grant entitled, "Assessing the Role of Covariates and Matching in Improving Bounds for Generalization and in Understanding Generalizations Over Time."

**Abstract**

Over the past ten years, propensity score methods have made an important contribution to improving generalizations from studies that do not select samples randomly from a population of inference. However, these methods require assumptions and recent work has considered the role of bounding approaches that provide a range of treatment impact estimates that are consistent with the observable data. An important limitation to bound estimates is that they can be uninformatively wide. This has motivated research on the use of propensity score stratification to narrow bounds. This article assesses the role of distributional overlap in propensity scores on the effectiveness of stratification to tighten bounds. Using the results of two simulation studies and two case studies, I evaluate the relationship between distributional overlap and precision gain and discuss the implications when propensity score stratification is used as a method to improve precision in the bounding framework.

**Introduction**

Randomized controlled trials (RCTs) are considered the strongest tool for assessing the causal impact of an intervention or program (Shadish, Cook, & Campbell, 2002). When treatments are randomly assigned in RCTs, this ensures an important aspect of the internal validity of the results; specifically, treatment randomization strengthens the extent to which a causal relationship exists between the treatment and outcome. With the proliferation of RCTs in fields such as education, social welfare, and medicine, policymakers and practitioners have grown increasingly interested in both the internal validity and *external validity*, or generalizability, of the studies' results to target populations of inference. In practice, the generalizability of a study's results is strongest when the study sample is randomly selected from the population of inference. However, random sampling is rare (Olsen, Orr, Bell & Stuart, 2013) and for the past ten years, statisticians have developed propensity score methods to improve generalizations from nonrandomly selected samples (Tipton, Hedges, Vaden-Kiernan, Borman, Sullivan, & Caverly, 2014; Tipton, 2013; Stuart, Cole, Bradshaw, & Leaf, 2011; Cole & Stuart, 2010).

Propensity score methods have made an important contribution to generalization research, but they depend on several assumptions, some of which are controversial in practice. For example, one assumption, sampling ignorability, requires that the propensity scores include all possible covariates that explain differences between the nonrandomly selected sample and the population of inference. When specific covariates are omitted or are not collected, sampling ignorability does not hold. Importantly, if sampling ignorability and other core assumptions are violated, then this has implications for policy decisions since evidence-based policymaking often relies on the validity of inferences from experimental studies. Recent work has considered the role of bounding approaches to derive alternative estimates of treatment impacts under fewer or alternative assumptions (Chan, 2017, 2019). In place of a single point estimate, bounds provide a range of plausible values of the treatment impact that are consistent with the observable data.

One important limitation to the bounding framework is that even with alternative assumptions, the bound estimates may be too wide to be substantively meaningful. Prior studies have shown that covariate and propensity score stratification are effective methods for tightening bounds, provided that the covariates satisfy certain conditions. For example, in the principal stratification framework, Miratrix, Furey, Feller, Grindal, & Page (2018) found that stratifying by covariates that were predictive of the outcome or of posttreatment behavior was effective in tightening bounds of treatment effects. Lee (2009) estimated bounds for the treatment effect of Job Corps on wages and found that stratification using covariates that predicted differences in wages was useful in narrowing bounds. In the generalization framework, Chan (2019) found that propensity score stratification narrowed bounds when the covariates were predictive of both sample selection and treatment effects. In these examples, stratification is used as a post hoc method to divide a selected sample of individuals based on a set of observable characteristics (covariates) such that individuals in the same subgroup are more alike than individuals in different subgroups. Stratification improves the precision of bounds (tightens them) by leveraging the similarity among subgroups.

An important question is how distributional overlap affects stratification as a method to tighten bounds. Broadly, in the generalization framework, distributional overlap refers to the proportion of population units whose range of propensity scores coincide with those in the sample (Tipton, 2013). Intuitively, if the distributional overlap between the sample and the population is weak, then stratification is potentially less effective in narrowing bounds, leaving open the question of whether stratification is a useful method to improve precision. The purpose of this article is to explore the relationship between distributional overlap in propensity scores and the precision gain attained under propensity score stratification in the bounding framework for generalization. Using two simulation studies and two case studies, I evaluate the impact of different values of distributional overlap on precision gain and identify the conditions under which stratification is most effective in tightening bounds. Because stratification is often used to improve precision in generalization research (Tipton, 2013; Tipton et al., 2014), the results of this study would inform evaluation practitioners of the factors that moderate the effectiveness of the method. Additionally, while the focus of this study is on generalization, the results are relevant for a potentially broader range of studies as stratification is commonly used to improve precision in parameter estimates (Lohr, 2009).

The article is organized as follows. First, I introduce the notation used throughout the study and discuss the assumptions required for propensity score methods. Second, I provide a brief summary of bounding methods for generalization. In the same section, I describe the role of propensity score stratification as a method to tighten bounds. Third, I formally define distributional overlap and discuss the results of two simulation studies that assess the effect of distributional overlap on the precision gains of bounds. I then turn to two case studies to analyze the relationship between distributional overlap and precision gain in practice. Finally, I conclude with a discussion of the implications of the use of propensity score stratification to improve precision in the bounding framework.

**Notation and Assumptions**

Consider a population $P$ of $N$ units of which a sample $S$ of $n$ units is selected into the experimental study. Units in this framework can refer to schools, clinics, or other aggregates of individuals. Throughout, I assume that the sample $S$ is not randomly selected from the population $P$, but that within the study, the units, $n$, are randomly assigned to each treatment condition. For each unit $i, i = 1, \ldots, N$, let $W_i$ be a binary treatment assignment indicator such that $W_i = 1$ if unit $i$ is randomly assigned to treatment and 0 otherwise. Let $Z_i$ be a sample selection indicator such that $Z_i = 1$ if unit $i$ is selected into the study sample and 0 otherwise. Lastly, I define $Y$ to be the outcome and $Y(1), Y(0)$ to be the potential outcomes of each unit under treatment and control, respectively (Rubin, 1977). The treatment effect for unit $i$ is defined by the difference in potential outcomes, $\Delta_i = Y_i(1) - Y_i(0)$.

To estimate $\Delta_i$, I assume that the stable unit treatment value assumption (SUTVA) holds for both the sample and population (Tipton, 2013; Rubin, 1978, 1980, 1986). Under SUTVA for the sample, the potential outcomes for unit $i$, $Y_i(1), Y_i(0)$, depend only on the treatment received by $i$ and not on the treatment received by unit $j$ for $i \neq j$. This assumption implies that the response to treatment among individuals in one unit does not depend on the treatment assigned to individuals in a different unit. Additionally, this assumption requires that there is only one version of the treatment. SUTVA for the population requires three assumptions. First, the conditions for treatment assignment must also hold for sample selection. Second, the potential outcomes do not depend on the proportion of units selected into the study. Third, there is no interference between units, both among the treatment and control groups and among the sampled and non-sampled groups.

If SUTVA holds for both the sample and the population, then the sample average treatment effect, $\Delta_{\text{SATE}}$ (SATE), is given by:

$$\Delta_{\text{SATE}} = E(\Delta_i | Z = 1) = E(Y_i(1) - Y_i(0) | Z = 1) \tag{1}$$

When treatment is randomly assigned, an unbiased estimator of $\Delta_{\text{SATE}}$ is given by $\hat{\Delta}_{\text{SATE}} = \frac{1}{n}\sum_{j \in \{Z=1\}} Y_j(1) - Y_j(0)$. In the generalization framework, the parameter of interest is the population average treatment effect, $\Delta_{\text{PATE}}$ (PATE), defined as:

$$\Delta_{\text{PATE}} = E(\Delta_i | Z = 1)\Pr(Z = 1) + E(\Delta_i | Z = 0)\Pr(Z = 0) \qquad (2)$$

Importantly, the SATE and PATE are equivalent only when treatment effect heterogeneity and sample selection are independent; namely, the two estimates are equivalent when the factors that explain variation in treatment effects among individuals are unrelated to the factors that affect participation in a study (Imai, King, & Stuart, 2008; Rubin, 1974). This condition holds when there is *no* treatment effect heterogeneity; namely, that treatment effects are constant across individuals in the sample and population (i.e., $(E(\Delta_i | Z = 1) = E(\Delta_i | Z = 0))$). This condition also holds when every unit is included in the study so that there is only one group (i.e., $\Pr(Z = 1) = 1$). Otherwise, an estimate $\hat{\Delta}_{\text{SATE}}$ may be unbiased for the SATE, but not necessarily for the PATE.

**Propensity Scores for Generalization**

When the sample $S$ is not a random subset of the population *P,* propensity score methods match or reweight the sample units *n* so that the resulting sample is compositionally "like" the target population. In generalization, propensity scores model the probability of selection into the sample based on observable covariates that predict sample selection and moderate differences in the treatment effect. For each unit in *P,* let **X** denote a vector of observable covariates, which may include continuous variables such as academic achievement scores or categorical variables such as urbanicity. The sampling propensity score, $s(\boldsymbol{X})$, is defined as:

$$s(\boldsymbol{X}) = \Pr(Z = 1 | \boldsymbol{X}) \qquad (3)$$

where $s(\boldsymbol{X})$ are the conditional probabilities of sample selection, conditional on the observable covariates $\boldsymbol{X}$. Propensity scores have the advantage of being balancing scores where units with similar propensity scores have similar distributions in the covariates $\boldsymbol{X}$ (Rosenbaum & Rubin, 1983). As a result, matching units in the sample and population by propensity scores is equivalent to matching by all the variables in $\boldsymbol{X}$. A common method of estimating the propensity scores $s(\boldsymbol{X})$ is with logistic regression with an intercept term $\alpha_0$:

$$\text{logit}\big(s(\boldsymbol{X})\big) = \log\left(\frac{s(\boldsymbol{X})}{1-s(\boldsymbol{X})}\right) = \alpha_0 + \alpha_1 X_1 + \cdots + \alpha_q X_q \tag{4}$$

based on $\boldsymbol{X} = (X_1, X_2, \dots, X_q)$ covariates.

**Assumptions for Propensity Scores**

Propensity score methods for generalization depend on several assumptions. In addition to SUTVA, the validity of propensity score-based estimates requires that treatment assignment be strongly ignorable (Stuart et al., 2011). This assumption is satisfied in evaluation studies with treatment randomization. In observational studies without treatment randomization, treatment assignment is considered strongly ignorable when researchers collect covariates that moderate treatment effect variation. In this case, treatment assignment is "as good as" random when all the covariates that affect heterogeneity in treatment effects are included in the analysis.

In addition to strong ignorability of treatment assignment, propensity score methods assume that *sampling ignorability* holds (Tipton, 2013). Sampling ignorability is specific to generalization and is met in studies where the sample is randomly selected from the population. In the absence of random selection, sampling ignorability holds when researchers collect information on variables that moderate both treatment effect variation and sample selection. Like strong ignorability of treatment assignment, sample selection is considered "as good as" random when the covariates (or a function of them such as propensity scores) that affect sample selection and treatment effect heterogeneity are included in the analysis. Formally, sampling ignorability assumes that:

$$\Delta = Y(1) - Y(0) \perp Z | s(\boldsymbol{X}) \tag{5}$$

$$\Delta = Y(1) - Y(0) \perp Z | s(\boldsymbol{X}) \text{ and } 0 < s(\boldsymbol{X}) \leq 1 \tag{6}$$

Equation 5 states that when the analysis includes the propensity scores $s(\boldsymbol{X})$, the treatment effect $\Delta$ is independent of sample selection. Specifically, when propensity scores are used, the treatment effect does not depend on whether a unit is selected into the study. Under (6), every unit in the population (sample) has a comparable unit in the sample (population). This is equivalent to the assumption that the distribution of covariates $\boldsymbol{X}$ in the sample and population share common support (Tipton, 2014).

**Bounding Methods for Generalization**

Among the assumptions, sampling ignorability is arguably the most controversial because it relies on conjectures on the relationship between the factors that affect sample selection and the factors that moderate treatment effects. Recent work has considered the role of bounding methods in generalization when core assumptions such as sampling ignorability are violated in practice (Chan, 2017). Bounding methods were first discussed in Manski (1990) who recommended that the statistical analysis start with the data alone so that researchers approach the problem from the same starting point. An important feature of bounding frameworks is that with fewer or alternative assumptions, researchers can estimate the $\Delta_{\text{PATE}}$ by a range of plausible values that are consistent with the observable data. For generalization, the goal of bounding approaches is to derive alternative estimates of the PATE without relying on assumptions such as sampling ignorability.

***Worst-Case Bounds***

To derive the bounds for the PATE, consider the simple case where the outcome Y is continuous but bounded in the same known range $[Y^L, Y^U]$ in both the sample and the population. Importantly, I assume that both SUTVA and strong ignorability of treatment assignment holds, but sampling ignorability may be violated. Additionally, I assume that the outcomes Y are measured without error in the sample and population. Then, using the observable data and given assumptions, the "worst-case bounds" for the $\Delta_{PATE}$ are given by:

$$\Delta_{PATE} \in [\Delta^L, \Delta^U] \tag{7}$$
$$\Delta^L = \Delta_{SATE} \Pr(Z = 1) + (Y^L - Y^U) \Pr(Z = 0)$$
$$\Delta^U = \Delta_{SATE} \Pr(Z = 1) + (Y^U - Y^L) \Pr(Z = 0)$$

The term $\Delta_{SATE} = E(Y(1)|W = 1, Z = 1) - E(Y(0)|W = 0, Z = 1)$ is the true SATE. To estimate the bounds in (7), we replace $\Delta_{SATE}$ with its estimate $\widehat{\Delta}_{SATE}$ (such as $\widehat{\Delta}_{SATE} = \frac{1}{n}\sum_{j \in \{Z=1\}} Y_j(1) - Y_j(0)$) and the probabilities $\Pr(Z = 1)$ and $\Pr(Z = 0)$ with the proportion of units selected and not selected into the sample, respectively. Note that the bounds in (7) are known as the "worst-case bounds" because they are the tightest bounds based solely on the data and knowledge of the range $[Y^L, Y^U]$. Details of the derivations of the bounds can be found in the Appendix.

***Narrowing Bounds with Propensity Score Stratification***

An important limitation to the bounds in (7) is that they are wide and never informative of the sign of the PATE (Chan, 2017). Prior work on bounding has shown that stratification can tighten bounds (improve precision) when the covariates used to stratify the sample satisfy certain conditions (Miratrix et al., 2018; Long and Hudgens, 2013). Chan (2019) found that propensity score stratification was effective in narrowing bounds when the covariates used in the propensity score model were predictive of sample selection and the outcome. Importantly, Chan found that even when the propensity score model included covariates that were only weakly predictive of sample selection or the outcome, stratification still yielded precision gains.

To estimate the bounds under propensity score stratification, let $P$ be divided into $k$ strata, each of which contains $N_k$ population units and $n_k$ sample units. Let $S_j$ define a unit's stratum membership where $S_j \in \{1,2,\dots,k\}$. For example, $S_j = 1$ implies that unit is a member of stratum 1. Within each stratum $k$, the PATE is defined as $\Delta_{\text{PATE}}{}^k = \sum_{j:S_j=k} \frac{1}{N_j}\big(Y_j(1) - Y_j(0)\big)$. The overall bound of the PATE is derived by averaging the lower and upper bounds across the strata:

$$\Delta_{\text{PATE}} \in [\Delta^L, \Delta^U] \tag{8}$$

$$\Delta^L = \sum_{j:S_j=1}^{k} \frac{N_j}{N}\big(\Delta_j{}^L\big)$$

$$\Delta^U = \sum_{j:S_j=1}^{k} \frac{N_j}{N}\big(\Delta_j{}^U\big)$$

We calculate the stratum-specific lower and upper bounds $\Delta_j{}^L, \Delta_j{}^U$ as follows. In stratum $j$, $j = 1,\dots k$, the lower bound is $\Delta_j{}^L = \Delta_{\text{SATE,j}}\Pr(Z=1)_j + (Y^L - Y^U)\Pr(Z=0)_j$ where $\Delta_{\text{SATE,j}}$ is the stratum-specific SATE estimated by $\hat{\Delta}_{\text{SATE,j}} = \frac{1}{n_j}\sum_{i\in\{Z=1\}} Y_i(1) - Y_i(0)$. The probabilities $\Pr(Z=1)_j, \Pr(Z=0)_j$ are estimated using the proportion of sample and non-sample units in stratum $j$, respectively. The upper bounds $\Delta_j{}^U$ are computed in an analogous way using the appropriate substitutions. In the stratified bounds, the term $N_j/N$ is the proportion of population units in stratum $j$, $j = 1,\dots k$, which serves as a stratum weight. When the population is stratified into equally sized strata, the stratum weight $N_j/N$ is equivalent to $1/k$. Although the population $P$ can be stratified in different ways, Cochran (1968) showed that stratification with $k = 5$ equally sized strata was optimal in terms of bias reduction and ease of implementation. However, it is important to note that five strata are not always possible, particularly when it leads to strata that have no sample units. In these cases, researchers should reduce the number of strata to ensure that each stratum contains enough sample units to estimate standard errors of treatment effects.

**When does stratification help?**

When comparing the standard (unstratified) bounds with the propensity score-narrowed (stratified) bounds, we see that the width of the bounds of the PATE are governed by three main terms: the $\Delta_{\text{SATE}}$, the likelihood of being in the study (Z) and the range between $Y^L$ and $Y^U$. This implies that one can tighten the bounds by changing one or more of these parameters. For example, if the probability $\Pr(Z = 1)$ increases or, conversely, if $\Pr(Z = 0)$ decreases, then this leads to tighter bounds. When $\Pr(Z = 1)$ increases, the sample size increases as a greater number of units are selected into the sample. Alternatively, another way to tighten the bounds is to reduce the range between $Y^L$ and $Y^U$. However, this method may be less flexible since the ranges of outcomes, $[Y^L, Y^U]$, are generally fixed.

Because the probabilities and range are not necessarily impacted by stratification, this suggests that stratification as a method to improve precision largely operates through the remaining term, $\Delta_{\text{SATE}}$. In practice, stratification improves precision by reducing the impact of variability in parameter estimates. For example, in survey sampling, stratification is used to divide a population into smaller subgroups (strata) such that individuals within strata are more alike than individuals across different strata (Lohr, 2009). By "alike," I refer to similarity based on a set of observable characteristics. In the survey sample context, stratification leverages the homogeneity within strata and reduces the variability due to sampling error to improve the precision of parameter estimates. In this study, because the focus is on generalizations from nonrandom experimental samples, the two main sources of variability are attributed to sample selection and treatment effect heterogeneity. As a result, when units in the sample and population are divided based on covariates (or functions of covariates) that moderate sample selection and variation in treatment effects, stratification can improve precision by strengthening the homogeneity based on the covariates.

**The Role of Distributional Overlap**

Current research has primarily focused on identifying the relevant covariates when evaluating the effectiveness of stratification to improve precision. An important question is how distributional overlap, referred to as overlap hereafter, affects propensity score stratification as a method to narrow bounds. Tipton (2013) found that when there is insufficient overlap between the sample and the population, this reduces the number of propensity score strata that can be created, which limits the amount of bias reduction that can be expected under stratification. The goal of this study is to address a similar concern by assessing how overlap affects the amount of precision gain that can be expected under propensity score stratification in the bounding framework for generalization.

The concept of "common support" in the sampling ignorability assumption of (5) and (6) has sometimes been used synonymously with the idea of overlap. However, in this framework, I distinguish between the two concepts by formally defining overlap in terms of the propensity scores. Let $s_S(X)$ and $s_P(X)$ be the propensity scores estimated among units in the sample and in the population, respectively. Overlap ($\Omega$) in the propensity scores is defined as:

$$\Omega = \int_{\min s_S(X)}^{\max s_S(X)} s_P(X) dp \tag{9}$$

which is the proportion of population units whose estimated propensity scores lie within the range of propensity scores of the sample. As a proportion, overlap is bounded between zero and one. Larger values of overlap (closer to 1) imply that a greater proportion of population units have similar propensity scores with the sample. Although the min $s_S(\boldsymbol{X})$ and max $s_S(\boldsymbol{X})$ can be based on the actual minimum and maximum propensity score in the sample, I follow the observational studies literature and define the min and max to be the 5th and 95th percentile of the propensity score distribution in the sample.

Prior work by Tipton et al. (2017) found that under random sampling, overlap depends on two factors: the number of covariates in the propensity score model $q$ and the sample size of the experimental study $n$. Specifically, if the number of covariates $q$ increases or the sample size $n$ decreases, then overlap decreases. For the former, overlap decreases because additional covariates increase the chance that some covariates will have different distributions in the sample and the population. Because the propensity scores are based on these covariates, including more covariates increase the chance that the estimated propensity scores between the sample and population will be different, which contributes to weaker overlap. With respect to sample size, smaller sample sizes are associated with more variability and this contributes to weaker overlap.

Although the analytic results of Tipton et al. (2017) do not apply to the nonrandom samples considered in this framework, overlap may still depend on the same parameters of $q$ and $n$. The purpose of this study is to assess the individual and collective impacts of these parameters, in addition to the role of covariates, on overlap and on the effectiveness of stratification to improve precision in bounds. If overlap in the propensity score distributions is weak, then it is possible that stratification may not be as effective in narrowing bounds, even if the covariates used are predictive of sample selection and the outcome. Additionally, when overlap is weak, another important question is whether the effects on precision gain can be "balanced" by changes in other parameters. Overall, addressing these concerns will have implications for the usefulness of stratification as a method to improve precision.

**Simulation Studies**

To explore the relationship between overlap and precision gain under propensity score stratification, I conducted two simulation studies. Both simulations are designed to mimic features of data sets found in many generalization studies; namely, the simulations are based on nonrandomly selected samples and the covariates represent a variety of variable types. Simulations 1 and 2 differed in terms of the outcome model. Simulation 1 represents a common framework where a set of covariates are predictive of both the selection and outcome variables. Simulation 2 represents an extreme case where the covariates that predict selection are different from those that predict the outcome. I include both simulations to assess the effect of covariates on overlap and precision gain. In the interest of brevity, I provide the details of the simulation setup in the Appendix.

**Simulation Parameters**

For Simulation 1 and 2, I varied two main parameters: the number of covariates used in the propensity score model $q$ and the sample size *n.* In the following sections, I provide specific details to describe how these parameters were varied.

### *Number of Covariates q in Propensity Score Model*

For both simulations, I estimated multiple sets of propensity scores by varying the number of covariates $q$ in the propensity score model. The smallest subset consisted of *q = 2* covariates (from six covariates $\boldsymbol{X} = X_1, X_2, X_3, X_4, X_5, X_6$) while the largest included all *q = 6* covariates, which led to 57 sets of propensity scores in total. Since multiple combinations of covariates led to similar estimated propensity scores and overlap, I used a subset of the 57 propensity score sets to study the effects of overlap on the precision gains under stratification.

***Sample Size n***

To assess the effect of sample size, I fixed the set of covariates $X$ used to estimate the propensity scores and used $X_1, X_2$ for Simulation 1 and $X_4, X_5$ for Simulation 2. I chose these covariates because they provided the largest overlap under the initial sample size of $n = 0.05N$. Thus, the analysis focuses on the effect of diminishing sample size on the maximum initial overlap. Although I fixed the set of covariates based on initial overlap, the analysis for sample size can also be done with different sets of covariates. Given the fixed set of covariates, I systematically reduced the sample size from the original $n = 0.05N$ to $n = 0.0055N$. The simulations stopped at $n = 0.0055N$ as further reductions were associated with little to no precision gain under propensity score stratification.

**Simulation Results**

This section describes the overall trends in the relationship between overlap and precision gain when the parameters of *q* and *n* are varied. For each simulation, I estimate two sets of bounds: the standard (unstratified) bounds using the whole sample and the stratified (propensity score narrowed) bounds by dividing the sample and population into equally sized strata using the estimated propensity scores. I define precision gain as the proportion reduction in bound width between the unstratified and stratified bounds. All results are based on 200 replications of each simulation study.

***Overlap and the Number of Covariates q***

Table 1 shows the relationship between overlap and precision gain when the number of covariates *q* used in the propensity score model changes. The results are based on a sample size $n = 0.05N$. For the purpose of illustration, I fixed the coefficients $\boldsymbol{\alpha} = (\alpha_1, \alpha_2, \alpha_3, \alpha_4, \alpha_5, \alpha_6) = (2, 2, 0.1, 0.1, 0.1)$ in the outcome models so that the average $R^2$ (which measures the predictiveness of the covariates in the outcome model) was approximately 0.90 in both simulations. For Simulation 2, all the covariate combinations included $X_4, X_5, X_6$.

Table 1 illustrates that as the number of covariates $q$ increases, the average overlap decreases, a result that is consistent with the random sampling case. In both simulations, when the propensity score model includes all six covariates $\mathbf{X}$, the average overlap is approximately 0.57, which implies that a little over half of the population units have propensity scores in the same range as the units in the sample. When the overlap $\Omega = 0.57$ with six covariates, the stratified bounds are 32.4% narrower in Simulation 1 and 27.3% narrower in Simulation 2, compared to the standard bounds. Both the overlap and the precision gain increase when the number of covariates in the propensity score model drops from six to two. For example, when $q$ = 2, the average overlap increases by 38% to 0.792 in Simulation 1 and by 54% to 0.882 in Simulation 2. With the improved overlap under two covariates, the precision gains are nearly double what they were when there were six covariates in the propensity score model.

TABLE 1

In Simulation 2, the covariates that predict the outcome are completely disjoint from those that predict selection. As a result, overlap is larger in Simulation 2, but because the propensity score model excludes covariates that are important for selection, the average precision gain is smaller. Table 1 illustrates that when stratification is used in this extreme case, the precision gains are about 20 – 30% smaller. For example, when the propensity score model only includes $q = 2$ covariates, the average precision gain is 0.481, which is 20% smaller compared to the gains seen in Simulation 1. However, when the number of covariates $q$ increases (and as a result, overlap decreases), the average difference in precision gains between the two simulation studies becomes smaller.

Although the precision gains are smaller in Simulation 2, overlap is higher because the study included covariates that were not predictive of selection. Since the included covariates do not explain differences between the sample and population, their distributions between the two groups were similar, and this led to higher overlap. However, as the results illustrate, even with the higher overlap and $R^2$ (from the outcome model), the average precision gain is more affected by the types of covariates included in the analysis. Specifically, when the propensity score model includes covariates that are only predictive of selection or the outcome, but not both, the average precision gains are consistently smaller, and the starkest differences are associated with the models that have few covariates ($q = 2$).

### *Overlap and Sample Size n*

Figures 1 and 2 illustrate the relationship between overlap and precision gain with various sample sizes *n*. The largest overlap in both figures was based on a sample size *n* that comprised 5% of the population while the smallest overlap was based on an *n* that was approximately 0.55% of the population. I incorporate the range of $R^2$ values in both figures to illustrate the trends in precision gain when the predictiveness of the covariates in the outcome model changes. To focus on the effects of sample size on overlap, I fixed the covariates used in the propensity score model to $\boldsymbol{X} = X_1, X_2$ for Simulation 1 and to $\boldsymbol{X} = X_4, X_5$ in Simulation 2. I chose these covariates because they yielded the highest initial overlap under $n = 0.05N$.

For Simulation 1, Figure 1 illustrates that when overlap is highest at 0.835, the maximum precision gain (corresponding to the highest $R^2$) is nearly 70%. As a result, when there is strong overlap and the covariates used in the propensity score model are highly predictive of the outcome, stratification shrinks bounds by over half of their original width. As the sample size $n$ decreases, the precision gain also decreases, and this decline is nearly constant across all values of $R^2$. For example, when overlap drops from 0.835 to 0.691, the average precision gain decreases from 0.70 to 0.55, an approximately 20% drop across each value of the $R^2$. When overlap is below 0.50, the maximum precision gain is only half the size of the gains seen when the overlap was highest at 0.835. Furthermore, when overlap is less than 0.50, the average $R^2$ for the covariates in the outcome model must be at least 0.25 to yield any meaningful precision gain. In other words, when overlap is weak, stratification can only narrow bounds if the covariates are moderately predictive of the outcome.

FIGURES 1, 2

The relationship among sample size, overlap and precision gain is similar for Simulation 2 when the covariates are only predictive of the outcome. Although the overlap between Figures 1 and 2 are not the same, their values were sufficiently similar to justify comparisons. From Figure 2, in the best-case scenario, when overlap is at its highest value of 0.882, stratification narrows bounds by nearly half their original width. Interestingly, when overlap drops from 0.882 to 0.782 (an 11% drop), there is not much difference in precision gain, particularly when $R^2$ is at least 0.75. This implies that when the covariates are strongly predictive of the outcome, the precision gains are practically the same, regardless of whether overlap is 0.70 or higher. When overlap drops to its lowest value of 0.502, the maximum precision gain is 0.25, which, like Simulation 1, is nearly half the gains seen compared to the case with the highest overlap. Additionally, when overlap is at its lowest, stratification is only effective in narrowing bounds when the covariates are sufficiently predictive with an $R^2$ that is at least 0.25.

**Discussion**

The simulation results highlight several important findings. First, stratification is a useful method to narrow bounds, but it is significantly more effective when there is strong overlap and the propensity score model includes highly predictive covariates. This is seen in the figures where an overlap close to 0.90 yielded bounds that were up to 70% narrower compared to the unstratified case. Notably, even when overlap is in the $0.70 - 0.80$ range, stratification can narrow bounds by as much as 50%, particularly when the covariates are strongly predictive of the outcome ($R^2 \geq 0.50$). This improvement in precision can potentially make bound estimates of the PATE more informative, especially when the unstratified bounds are too wide to be useful. Second, even when overlap is only 0.50, stratification still yields precision gains if the covariates in the propensity score model are sufficiently predictive of the outcome. In the simulation framework, "sufficiently predictive" was associated with an $R^2$ that was at least 0.25. Additionally, the results of both simulations illustrate that different combinations of overlap and $R^2$ values can yield similar precision gains under stratification. For example, a study with a lower overlap of 0.691 had similar precision gains as a study with a higher overlap of 0.746 when the average $R^2$ of the covariates is approximately 0.50. Third, in the extreme case when propensity scores are estimated using covariates that are only predictive of the outcome (Simulation 2), overlap is larger, but the average precision gains are smaller, by as much as 30%. Furthermore, the results from Figure 2 suggest that in the extreme case of Simulation 2, when the covariates are already strongly predictive ($R^2 \geq 0.75$), increasing overlap does not have much of an effect on precision gain. Although the results from Simulation 2 represent an extreme case, they may still provide insight into the kinds of gains that can be expected under propensity score stratification, particularly if the covariates used are only weakly predictive of selection. Finally, the results of both simulations suggest that drops in overlap can result in even steeper drops in precision gain.

In the extreme case when overlap decreased from 0.835 to 0.494, this 30% decrease was associated with a 50% drop in the precision gain.

**Two Case Studies**

To evaluate the relationship between overlap and precision gain under propensity score stratification in practice, I turn to two case studies. Although both studies are educational examples, the relationship between overlap and precision gain is an important consideration for any study in which stratification is used to improve precision in treatment effect estimates. The case studies in this section are cluster randomized trials (CRTs) in which schools were assigned to a treatment or control condition. In both CRTs, the original sample of schools was not a random sample from a target population of inference, but within the sample, schools were randomly assigned to each experimental condition. Importantly, the estimated overlap in propensity score distributions differed between the studies, primarily due to differences in sample size. In this section, I perform a re-analysis of each example, describe the relationship between the estimated overlap and precision gain, and discuss any connections between the results and the simulation studies.

**Case Study 1: Indiana CRT**

From 2009 – 2010, the Indiana Department of Education and the Indiana State Board of Education implemented a new assessment system to measure annual student growth (Konstantopoulos et al., 2013). A sample of 54 K-8 (elementary to middle) schools volunteered to participate in the CRT, of which 33 were randomly assigned to use the state's assessment system (treatment) and the remaining 21 were assigned to control. In the treatment schools, students were given four diagnostic assessments that were aligned with the Indiana state exam and their teachers used information from the reports on the students' performance to dynamically guide their instruction. The effectiveness of the assessment system was measured using the Indiana Statewide Testing for Educational Progress-Plus (ISTEP+) scores in English Language Arts (ELA) and mathematics. The PATE is defined as the average difference in ISTEP+ scores between schools that used the assessment system and schools that did not. In the original study, the authors found an average treatment impact of 0.078 (SE = 0.050) and 0.127 (SE = 0.069) for ELA and mathematics, respectively, neither of which was statistically significant (Konstantopoulos et al., 2013).

Chan (2017) and Tipton et al. (2017) analyzed the Indiana study from a generalization framework using a population of $P = 1,514$ K-8 schools in Indiana during the 2009 – 2010 academic year. Because the sample of 54 schools was not a random subset of the population, both studies used propensity scores to match the CRT and Indiana population schools on 14 covariates to derive bias reduced estimates of the PATE. Table 2 provides the means of these covariates, which were taken from the Common Core of Data (National Center for Education Statistics; https://nces.ed.gov/ccd) and the Indiana Department of Education.

TABLE 2

Using the same set of 14 covariates, I estimated the propensity scores of selection into the Indiana CRT and stratified the combined sample and population schools into equally sized strata. Because the sample comprised 3.6% ( $54/1514 \approx 0.036$) of the population, the data only permitted three equally sized propensity score strata. Within each stratum, I estimated the worst-case bounds using the standardized ISTEP+ scores and compared the final stratified bounds with the unstratified estimates to estimate the precision gain. The bounds for the ISTEP+ scores use the range $[Y^L, Y^U] = [-1.37, 1.68]$ for ELA and $[Y^L, Y^U] = [-2.37, 1.99]$ for math, both of which are based on the standardized scores. The first row of Table 3 provides the bound estimates for ELA and mathematics in the Indiana CRT using the propensity score model with 14 covariates.

TABLE 3

For the Indiana CRT, the average overlap was 0.542, which implies that about 54.2% of the Indiana population schools had estimated propensity scores within the range of the sample. Compared to the simulation results, this study's overlap is on the lower end of the range, which is likely due to the small sample size. However, the high $R^2$ values (0.764 for ELA and 0.634 for math) illustrate that the 14 covariates are strongly predictive of the outcomes in both subjects. As a result, propensity score stratification tightened the bounds for ELA and mathematics by 39.1% and 27.7%, respectively.

Compared to the simulation results, the gains seen in the Indiana CRT are larger than those observed in the simulations. Although the Indiana CRT overlap of 0.542 was not included in the simulations, the precision gains seen in the Indiana study are close to the gains associated with similar overlap values in the simulations. Specifically, when the $R^2$ is nearly 0.75, the average precision gain in the simulations were between 0.25 and 0.33. These values are close to the 0.391 seen for ELA in the Indiana CRT. Similar results can be observed in the comparison for math.

**Case Study 2: SimCalc**

SimCalc is a mathematics software program that uses computer animations to teach concepts of rates and proportions. To assess the impact of this program on mathematics achievement, the research firm SRI International implemented two cluster randomized experiments, one of which was a pilot study, on a combined sample of 92 middle schools in Texas from 2008 – 2009 (Roschelle et al., 2010). In the original study, the principal investigators found a statistically significant main effect of 1.438 (SE = 0.143, p < 0.001) in students' gain scores in mathematics, which implied that students using SimCalc experience larger gains than students who did not receive the software. This effect is standardized in relation to the between-school variance.

Because the sample of 92 middle schools was not randomly selected from the population of 1,713 Texas middle schools in the given academic year, Chan (2019), O'Muircheartaigh & Hedges (2014) and Tipton (2013) assessed the generalizability of the SimCalc findings using propensity score methods. In these studies, propensity scores were used to match the SimCalc and Texas population schools on 26 covariates deemed relevant in predicting sample selection and the treatment impact of SimCalc. Table 4 provides the covariate means, the data of which are taken from the Texas Academic Excellence Indicator System (AEIS).

TABLE 4

Like the Indiana CRT, I re-analyzed the SimCalc study and computed propensity scores using the same 26 covariates from Table 4. Unlike the Indiana CRT, the SimCalc study allowed for five propensity score strata. The first row of Table 5 provides the estimated bounds, overlap, and precision gain for the PATE, which is defined as the difference in gain scores between the treatment and control schools. The bounds were estimated using the range $[Y^L, Y^U] = [-1.96, 3.24]$ for the standardized gain scores in mathematics.

TABLE 5

Table 5 shows that the average overlap in SimCalc is 0.74, which implies that 74% of the schools in the population had estimated propensity scores within the range of the propensity scores of the SimCalc sample. Note that the sample size of the SimCalc study is larger ($n = 92/1713 \approx 0.05$), which may partly explain the relatively higher overlap. However, the 26 covariates in SimCalc are notably weaker in predicting the outcome, with an average $R^2$ of less than 0.30. With an overlap of 0.74 and $R^2 = 0.279$, propensity score stratification narrowed the bounds for the PATE by 44.9%.

***Comparison of Case Studies***

As a final analysis, I compared the results of both case studies to assess whether the empirical results supported the simulation findings. Because the simulations suggest that overlap is affected by the number of covariates $q$, I selected a subset of five covariates from each study to estimate the propensity scores. These covariates collectively explained 10% of the variance in the respective outcomes ($R^2 \approx 0.10$). The decision to use five covariates was made for the sole purpose of comparison and the specific covariates were chosen to ensure that the average $R^2$ was comparable between studies. The purpose of this final analysis was to assess the effect of overlap on precision gain, with all else being nearly equal. The results can be found in the second rows of Tables 3 and 5.

Not surprisingly, with fewer covariates, the average overlap increased in both studies to 0.69 for the Indiana CRT and to 0.85 for SimCalc. However, the predictiveness of the covariates ($R^2 \approx 0.10$) is much smaller compared to the original propensity score models, particularly for the Indiana CRT. Between the two studies, the average gain was higher in SimCalc under the new propensity score model, which is consistent with the simulation results. Interestingly, for both studies, despite the decreases in $R^2$, the precision gains under the new model are nearly identical to those in the original propensity score models. This may be a result of the "counteracting" effect of stronger overlap. Note that while reducing the number of covariates improved overlap for both studies, the average overlap for the Indiana CRT (0.69) was still on the lower end, which again may be a consequence of the smaller sample size.

**Implications of Case Studies**

Overall, the results from the case studies are consistent with those found in the simulations and they suggest two main implications. First, when overlap is weak, the strength of the covariates' predictive power has a strong impact on the extent to which stratification can narrow bounds. This was seen in the Indiana study where, despite a weak overlap, the precision gains were comparable to SimCalc because of the stronger covariates. Second, while the number of covariates in the propensity score model is important, sample size may have a larger impact on overlap. This was seen in the head-to-head comparison between the studies where overlap increased in both cases, but it was still on the lower end for the Indiana CRT, likely due to the smaller sample size. As a result, focusing on the strength of the covariates may be especially important for smaller generalization studies where the average overlap may be weak and the extent to which stratification yields precision gains will necessarily depend on other factors.

**Discussion**

This article explores how overlap affects the extent to which propensity score stratification improves precision for bounds in generalization studies. Because the concepts of stratification are foundational to experimental design, it is important for researchers seeking to generalize from experimental studies to understand the factors that impact its effectiveness in practice. In this section, I conclude with three implications of this study and discuss their importance for evaluation practitioners and researchers.

1. **Stratification improves precision, but the magnitude of the gains depends on overlap and the covariates.** The results of this study illustrate that stratification is an effective method to improve precision, but it is most effective when both the overlap and covariates' predictive power are strong. This is especially pertinent for researchers generalizing from small studies (e.g., when $n/N < 0.05$) where the smaller sample sizes directly limit the maximum possible overlap. For these studies, strengthening factors such as the covariates' predictive power will be important to balance the limiting effects of weaker overlap.

2. **Overlap matters, but covariates may matter more.** The simulation studies highlighted that even when overlap is strong (above 0.70), stratification yielded meaningful precision gains only when the average $R^2$ was at least 0.25. As shown in the case studies, even when overlap was weak, its effects on precision gain were more than offset by stronger covariates. In the generalization framework, covariates play an important role in explaining variability due to sample selection and treatment effect variation. Because the effectiveness of stratification (by covariates or by propensity scores) depends on minimizing this variability, identifying and including the variables that explain the most variation may make a difference when using stratification to improve precision, particularly in smaller studies. Thus far, the empirical evidence on the types of covariates to include in the propensity score model remains limited. In practice, researchers should choose variables based on substantive theory for what might moderate the sample selection mechanism and treatment effect heterogeneity.

Considering the role of covariates, it is important to address the somewhat conflicting results presented in Table 1 that illustrated a tradeoff between overlap and model complexity. Specifically, when the propensity score model is based on fewer covariates, this was associated with larger overlap and precision gains for bounds. This raises the question of how researchers should consider the individual and collective impacts of model complexity (number of covariates), covariates, and overlap with the goal of attaining the largest precision gain under stratification. In light of the observations from the simulation and empirical studies, I argue that researchers should prioritize the *inclusion* of covariates, particularly ones that are strongly predictive of selection and treatment effects, for two reasons. First, the largest precision gains are consistently associated with the highest $R^2$ values so that the predictive power of the covariates is a critical factor, even when overlap is weak. Second, covariates play an important role in improving precision, but also in reducing *bias*. In the context of this study, the bias is due to the nonrandom nature of the sample selection process where units that self-select into the sample may be systematically different from those that do not participate in a study. Because the extent of bias reduction depends on the types of covariates included, researchers should opt for a bias robust approach where more covariates, not fewer, should be included for stratification. Thus, while increasing model complexity (more covariates) may be associated with less overlap, its effects may be counteracted by the gains in precision and bias.

3. **Omitted covariates also matter.** Propensity score-based methods, including stratification, assume that all relevant moderators of sample selection and treatment effect heterogeneity have been included in the analysis. However, this assumption is difficult to test and so researchers can never be certain if a study has omitted important variables. The challenge with omitted covariates is that estimates of treatment effects, both point estimates and bounds, are not necessarily unbiased and the extent of the bias depends on the role that the omitted variables play in satisfying the assumptions for propensity score methods.

   The bias from omitted covariates also affects the results of this study. Because the propensity scores are estimated with the observable covariates, the findings on the relationship between overlap and precision gain are limited by the strength of these variables alone to explain variability. Thus, the results of this study address the effect of overlap on precision and variance, but they do not account for the effect on bias. However, the simulations provided some insight into the effect of omitted covariates in Simulation 2 when several important moderators were excluded, and this exclusion limited the maximum amount of precision gain expected under stratification. Given the challenges of ascertaining whether important covariates have been omitted in practice, researchers may choose to conduct additional sensitivity analyses to assess the potential effect of omitted variables. While sensitivity analyses are not a solution to omitted covariate bias, they provide insight into the possible implications on parameter estimates.

In conclusion, the extent to which stratification is a useful method for improving precision is affected by the collective impacts of overlap, covariates and sample size. The findings of this study suggest that a general good practice in applications of stratification to improve external validity (generalizability) is to prioritize the inclusion of relevant covariates that moderate the mechanisms behind sample selection and variation in treatment effects. In the ideal case, covariates strengthen the effects of strong overlap in producing large precision gains. Importantly, in cases where overlap is weak, stratification with strong covariates can still yield moderately large precision gains.

Table 1. Relationship Between Overlap and the Number of Covariates $q$ in Propensity Score Models

| Number of Covariates $q$ | Simulation 1 | | Simulation 2 | |
|---|---|---|---|---|
| | Average Overlap $\widehat{\Omega}$ | Average Precision Gain | Average Overlap $\widehat{\Omega}$ | Average Precision Gain |
| 6 | 0.574 | 0.324 | 0.571 | 0.273 |
| 5 | 0.625 | 0.384 | 0.652 | 0.271 |
| 4 | 0.678 | 0.451 | 0.727 | 0.297 |
| 3 | 0.730 | 0.520 | 0.799 | 0.348 |
| 2 | 0.792 | 0.605 | 0.882 | 0.481 |

The coefficients in the outcome model $\boldsymbol{\alpha} = (\alpha_1, \alpha_2, \alpha_3, \alpha_4, \alpha_5, \alpha_6)$ were fixed to $(2, 2, 0.1, 0.1, 0.1, 0.1)$ so that the average $R^2$ in both simulations were approximately 0.90 in the outcome model. In Simulation 1, both the selection and the outcome model are based on $\boldsymbol{X} = X_1, X_2, X_3$. In Simulation 2, the selection model is based on $\boldsymbol{X} = X_1, X_2, X_3$, but the outcome model is based on $\boldsymbol{X} = X_4, X_5, X_6$.

Table 2. Covariate Means for Indiana CRT and Indiana Population Schools

| Covariate Descriptions | Indiana CRT $n = 54$ | Population $N = 1,514$ |
| --- | --- | --- |
| 2008 - 2009 ELA Test Scores | 19.4 | 18.8 |
| 2008 - 2009 Math Test Scores | 16.5 | 16.5 |
| 2009 - 2009 Attendance | 96 | 96 |
| 2008 - 2009 Full Time Staff | 24 | 28 |
| 2008 - 2009 Population of Students | 423 | 481 |
| 2008 - 2009 Pupil-Teacher Ratio | 18 | 17 |
| County Population | 109879 | 230487 |
| 2008 - 2009 Title I Status (%) | 80.0 | 80.0 |
| 2008 - 2009 Schoolwide Title I Status (%) | 50.0 | 50.0 |
| Proportion of Male Students | 50.0 | 50.0 |
| Proportion of White Students | 80.0 | 80.0 |
| Proportion of Special Education Students | 20.0 | 20.0 |
| Proportion of Free/Reduced Price Students | 50.0 | 50.0 |
| Proportion of Limited English Proficiency Students | 0.0 | 0.0 |

The covariate means are averaged at the school level in both the CRT and the population.

Table 3. Overlap and Precision Gain for Indiana CRT

| Number of Covariates | Subject | Unstratified Bounds | Stratified Bounds | $R^2$ | Overlap | Precision Gain |
|---|---|---|---|---|---|---|
| | ELA | [-2.93, 2.94] | [-1.78, 1.80] | 0.76 | | 0.39 |
| 14 | Mathematics | [-4.20, 4.21] | [-3.03, 3.05] | 0.63 | 0.54 | 0.28 |
| | ELA | [-2.93, 2.94] | [-1.90, 1.92] | 0.14 | | 0.35 |
| 5 | Mathematics | [-4.20, 4.21] | [-2.99, 3.01] | 0.08 | 0.69 | 0.29 |

The number of covariates refers to the variables included in the propensity score model. The unstratified bounds are the standard bounds. The stratified bounds are the propensity score-narrowed bounds. The $R^2$ measures the predictive power of the covariates in the outcome model.

Table 4. Covariate Means for SimCalc and Texas Population schools

| Covariate Descriptions | SimCalc<br>*n = 92* | Population<br>*N = 1,713* |
|---|---|---|
| Teacher tenure (mean) | 7.0 | 7.1 |
| Teacher experience (mean | 11.0 | 11.6 |
| Teacher-student ratio | 13.3 | 12.7 |
| % of African-American teachers | 4.9 | 8.4 |
| % of Hispanic teachers | 21.7 | 14.7 |
| Total number of teachers | 46 | 40 |
| % of teachers in 1$^{st}$ year of teaching | 8.5 | 8.3 |
| % of teachers with 1 – 5 years of experience | 30.1 | 28.0 |
| % of teachers with > 20 years of experience | 18.1 | 20.3 |
| % of students in disciplinary alternative education programmes | 3.6 | 3.1 |
| 7$^{th}$ grade retention rate | 1.3 | 1.8 |
| % of students who are mobile | 15.2 | 19.2 |
| % of 7$^{th}$ grade students | 36.1 | 31.2 |
| Total number of 7$^{th}$ grade students | 241 | 190 |
| % of African-American students | 7.4 | 11.8 |
| % of Hispanic students | 47.8 | 40.3 |
| % of students who are limited English proficient | 10.2 | 7.5 |
| % of economically disadvantaged students | 53.9 | 53.6 |
| % of students at risk | 41.1 | 43.5 |
| Students who are proficient in 7$^{th}$ grade reading (%) | 86.1 | 81.9 |
| Students who are proficient in 7$^{th}$ grade mathematics (%) | 75.6 | 72.8 |
| Students who are proficient in 3 – 11 mathematics (%) | 74.7 | 73.6 |
| Students who are proficient in 3 – 11 all subjects (%) | 63.4 | 63.3 |
| % of students with commended performance, grades 3 – 11, mathematics | 20.1 | 19.6 |
| % of students with commended performance, grades 3 – 11, reading | 8.7 | 8.7 |
| County of schools is rural (%) | 30.0 | 30.0 |

The covariate means are averaged to the school level for both SimCalc and the population.

Table 5. Overlap and Precision Gain for SimCalc

| Number of covariates | Unstratified Bounds | Stratified Bounds | $R^2$ | Overlap | Precision Gain |
|---|---|---|---|---|---|
| 26 | [-4.82, 4.99] | [-2.62, 2.78] | 0.28 | 0.74 | 0.45 |
| 5 | [-4.82, 4.99] | [-2.69, 2.84] | 0.11 | 0.85 | 0.44 |

The number of covariates refers to the variables included in the propensity score model. The unstratified bounds are the standard bounds. The stratified bounds are the propensity score narrowed bounds. The $R^2$ measures the predictive power of the covariates in the outcome model.

Figure 1. Relationship Between Precision Gain and Sample Size *n* (Simulation 1)

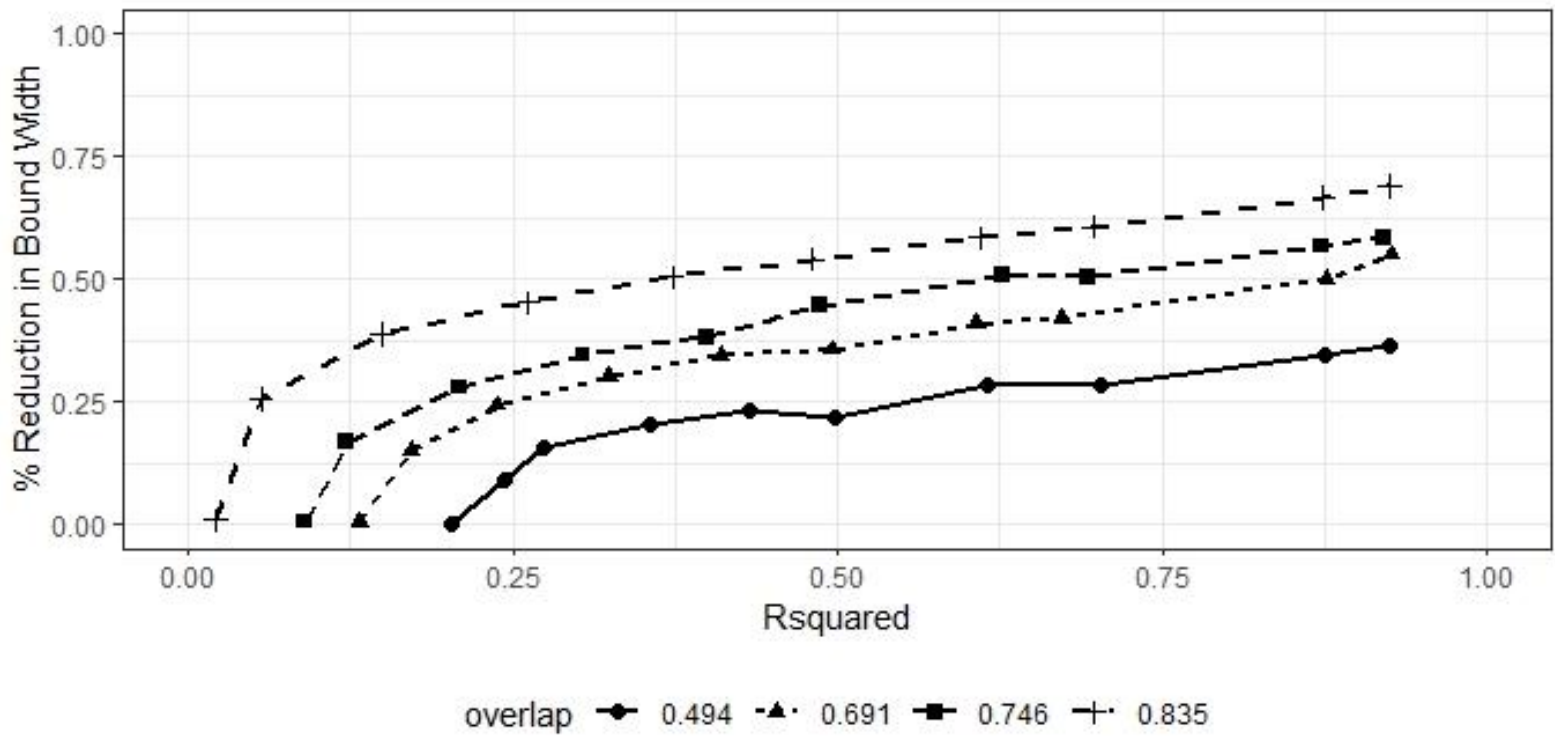

Note: Rsquared refers to the predictive power of the covariates in the outcome model. The % reduction in bound width is the difference in width between the standard (unstratified) bounds and the stratified bounds. For each overlap value, the sample sizes were as follows: $\Omega = 0.835$ ($n = 0.05N$), $\Omega = 0.746$ ($n = 0.011N$), $\Omega = 0.691$ ($n = 0.008N$), $\Omega = 0.494$ ($n = 0.0055N$)

Figure 2. Relationship Between Precision Gain and Sample Size *n* (Simulation 2)

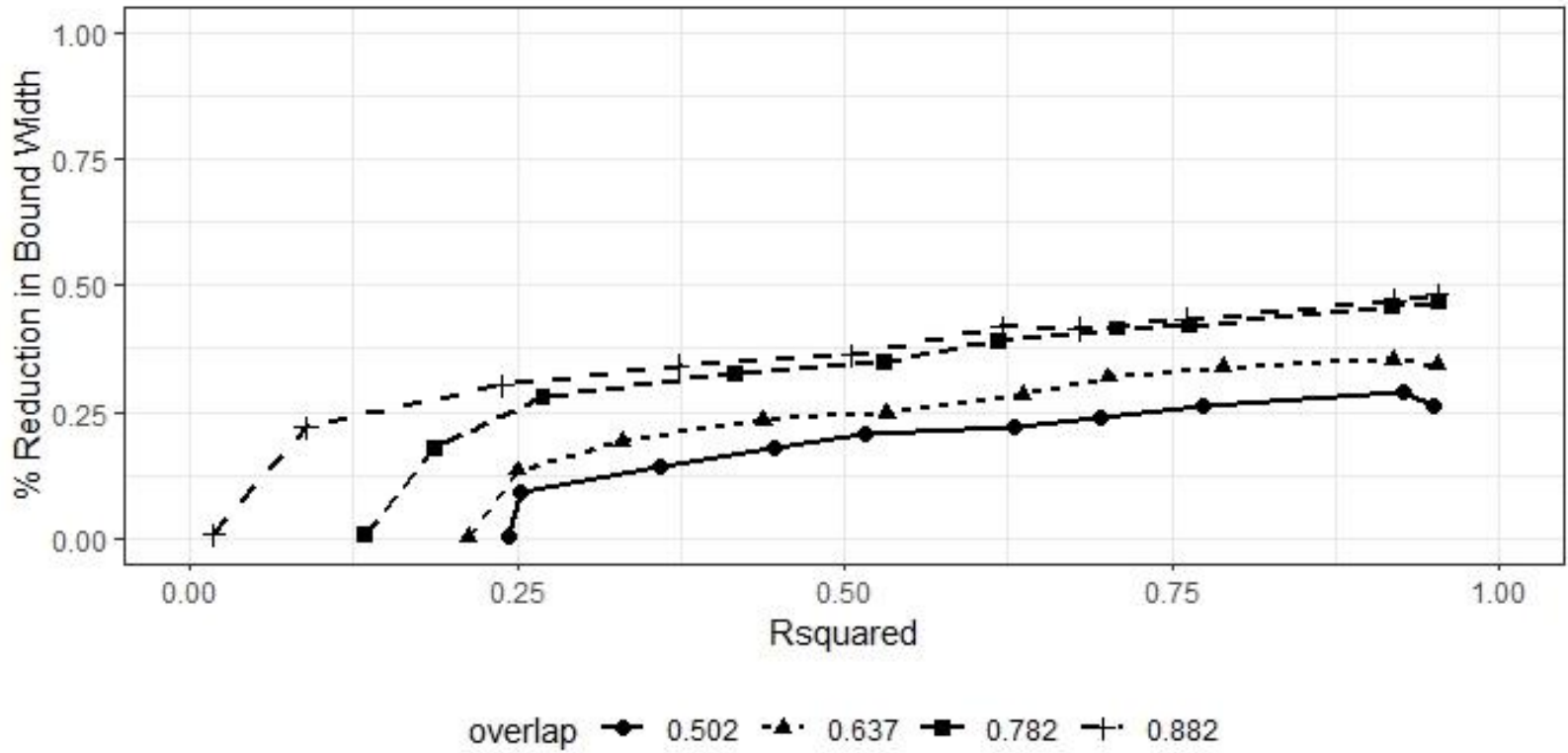

Note: Rsquared refers to the predictive power of the covariates in the outcome model. The % reduction in bound width is the difference in width between the standard (unstratified) bounds and the stratified bounds. For each overlap value, the sample sizes were as follows: $\Omega = 0.882$ ($n = 0.05N$), $\Omega = 0.782$ ($n = 0.01N$), $\Omega = 0.637$ ($n = 0.008N$), $\Omega = 0.502$ ($n = 0.005N$)